# High-speed CMOS compatible plasmonic modulator for non-contact wafer-level testing

Maryam Sadat Amiri Naeini,[1, 2] Pierre Berini[1, 2, 3*]

[1]Department of Physics, University of Ottawa, 150 Louis-Pasteur Pvt., Ottawa, Canada

[2]Nexus for Quantum Technologies Institute - NEXQT, University of Ottawa, 25 Templeton St., Ottawa, Canada

[3]School of Electrical Engineering and Computer Science, University of Ottawa, 800 King Edward Ave., Ottawa, Canada

*Corresponding author email: pberini@uottawa.ca

**Abstract**

Wafer-level testing is an important step for process and quality control of electronic chips in integrated circuit (IC) manufacturing which occurs before packaging. The process of wafer probing in its conventional contacting schemes, becomes more complicated as ICs move to smaller technology nodes and more compact designs, greatly increasing testing costs. Non-contact optical wafer probing can overcome physical probing complications, reducing costs, and increasing throughput and reliability. In this paper, a CMOS compatible, broadband (22 GHz), small footprint (5 µm dia.) plasmonic electro-optic modulator of low insertion loss (4 dB) and wide optical working bandwidth (100 nm) is proposed and demonstrated as a potential solution for wafer-level optical testing. The device modulates in reflection an incident optical carrier emerging from an optical fiber in a non-contact arrangement, to work as a data output channel from the wafer. A modulation depth of over 2% is achieved which should be sufficient to meet the requirements of wafer-level testing. The device can be placed anywhere on wafer.

**Introduction**

In the ever-growing landscape of semiconductor electronics, the demand for smaller, faster, more efficient and more powerful electronic devices has led to continuous evolution in design and manufacturing processes. These innovations apply increasing pressure on the crucial step of wafer testing (probing) [1] [2] [3] [4] [5] [6] [7] [8], which serves to assess the performance of high-volume, high-density, and high-value Si CMOS ICs (complementary metal oxide semiconductor integrated circuits) such as microprocessors, memories, and systems-on-a-chip. Wafer testing occurs after wafer manufacturing and test data are used for yield estimates, process quality control and product verification before die singulation and packaging – wafer test is a gateway to completed and qualified electronic products.

Traditional wafer probing methods involve physical engagement between hundreds of needles in probe cards with electrical pads on wafer, a process laden with a variety of challenges, including the manufacturing of customized probe cards which is complicated and costly, damage and material accumulation on probe needles after several touchdowns, wafer and device damage due to probe contact, limited bandwidth of test signals, electromagnetic interference and impedance mismatch, low test throughput, and limited scalability. Non-contact wafer probing would be paradigm-shifting, motivating the investigation of a variety of innovative approaches involving e-beam [9] [10], plasmas [9], ultra-violet lasers

[11] and other optical interrogation methods [12] [13], capacitive coupling to an AFM (atomic force microscope) probe [14] [15] [16] [17], and broadside-coupled striplines [18], all of which are only used for defect inspection at small scales and mostly aren't suitable for volume testing. Near field radio frequency interrogation by inductive coupling has also been proposed [11] but this approach suffers from limitations in scalability due to crosstalk, and difficulty in maintaining the small distance required between the wafer and the test head. Considering the challenges posed by various methods, optical solutions are promising if they can be integrated into industrial electronics manufacturing processes [12]. However, more work is required involving new designs and techniques, especially for system-in-package (SiP) and 3D IC testing. The primary criteria for a successful practical solution include three key factors: (*i*) optoelectronic performance that is sufficient to establish communication between electronics on-wafer and the test instrument, (*ii*) optoelectronic device compatibility with CMOS fabrication processes [19], and (*iii*) optoelectronic device dimensions that are of the same size (or smaller) than the test pads on chip that would be replaced. In this context, CMOS compatible metal-oxide-semiconductor (MOS) based plasmonic optoelectronic devices, such as electro-optic modulators and photodetectors hold the promise of a potential solution. In such a scheme, a photodetector on chip would be used to receive an optical carrier modulated with test instructions and data, and a modulator on chip would be used to modulate an incident CW (continuous-wave) optical carrier to send test data off-chip. The latter could be achieved using reflection modulators and photodetectors that can be probed by a fiber array probe card, as suggested in Fig. 1. The compact nature of plasmonic elements, enabled by the strong confinement of surface plasmon polaritons (SPPs), allows for their dimensions to be very small, *e.g.*, of the size or smaller than a diffraction-limited incident beam, which ensures minimal consumption of valuable area on wafer.

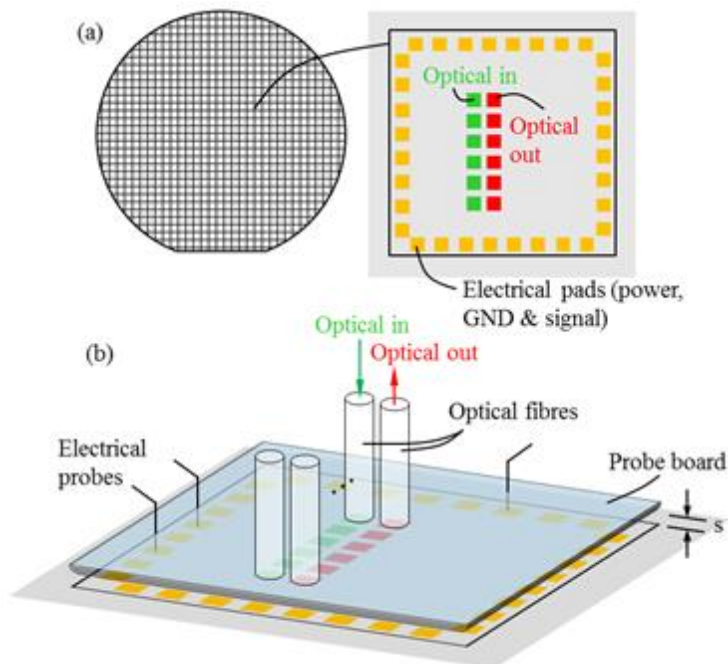

**Figure 1.** (a) Sketch of a Si wafer and of a die in expanded view showing power, ground and signal pads on the periphery and optical inputs (photodetectors – green squares) and outputs (modulators – red squares) in the central region. (b) Non-contact optical interrogation of a die under test using an array of fibers mounted to a probe card aligned to photodetectors and modulators on chip. The photodetectors receive a data stream on a modulated optical carrier (green), whereas the modulators receive a CW beam and return by modulated reflection a data stream on the same beam (red).

There have been substantial study and development of silicon [20] and silicon MOS based modulators [21]. Due to the indirect bandgap of Si and the lack of a linear electro-optic effect, the most common modulation approaches have involved interference or ring-resonator designs [20] [21] [22], exploiting the carrier refraction effect [23]. Interferometers have large footprints and ring resonators have high sensitivity to temperature and require tight fabrication tolerances, precluding the use of such structures for testing applications. Recently, plasmonic and hybrid-plasmonic modulators have also received significant attention due to the enhanced light-matter interaction which enables small active areas [24] [25]. The enhancement is due to the tight light confinement within the active area. However, such designs need to be integrated with silicon waveguides resulting in enlarged final device dimensions [22] [24] [25] [26] [27]. Furthermore, the light coupling methods in these devices are not well-suited to wafer probing purposes where the optical inputs and outputs should be accessible via the top surface of the wafer (*cf.*, Fig. 1), and the materials used for the active region are not part of CMOS material sets [28] [29]. These properties make existing plasmonic modulator concepts unsuitable for non-contact wafer probing applications.

Building on previous work carried out at low frequencies [30], we introduce in this paper a high-speed MOS-based plasmonic reflection modulator on Si that has the electro-optic bandwidth, modulation depth, optical insertion loss, and a coupling method suitable for wafer testing applications. Their compact dimension (down to 5 μm in diameter) enables operation up to 22 GHz as lumped elements, and in principle, can be integrated anywhere on wafer while consuming very little area. Combining the modulators with plasmonic photodetectors, based on tunnelling in a MOS structure [31], or on internal photoemission in a Schottky structure [32], could lead to a full non-contact optical test solution for electronic wafers following the approach of Fig. 1.

**Device structure and concept**

As shown in Fig. 2, the reflection modulator consists of a MOS capacitor bearing a plasmonic grating as the top metal contact. The grating enables plasmon excitation along the MOS structure upon optical illumination at perpendicular incidence at the grating design wavelength and with a polarization aligned perpendicular to the grating fingers. The MOS device is biased around the flat band voltage and driven between the accumulation and depletion regimes by applying an AC drive voltage. The AC voltage modulates the density of accumulated carriers in the semiconductor near the oxide layer, which modulates the effective index of the surface plasmons via the carrier refraction effect [21], thereby modulating the fraction of incident light that is coupled to the surface plasmons [30]. As a result, the light intensity reflected by the modulator is modulated by the applied signal.

Modulator chips were designed as circular parallel plate MOS capacitors of various diameters (5, 11, 17, 22 and 28 μm) formed as a t = 20 nm thick gold layer on 5 nm thick hafnium dioxide ($HfO_2$) layer on a heavily doped p-Si (ρ = 0.001-0.005 Ω cm) substrate. The gold layer was covered by a plasmonic grating of thickness H = 80 nm and of pitch selected in the range of Λ = 420 - 480 nm, following the momentum conservation condition at the working wavelength for the case of a normally incident Gaussian beam (Λ = $\lambda_0/n_{eff}$), where $n_{eff}$ is the effective refractive index of plasmons propagating along the bottom surface of the gold patch (along the MOS structure), and $\lambda_0$ is the free-space working wavelength ($\lambda_0$ = 1470 - 1680 nm). The chips were then passivated by a 554 nm SU8 protective layer and electrical contact pads were fabricated on this layer connected to the gold and silicon layers through metallized vias. The pitch of the pads was selected to accommodate 40 GHz ground-signal-ground (coplanar waveguide) probes. Further details on the fabrication of the modulators can be found elsewhere [33]. The schematic cross-section of the modulator in Fig. 2(a) illustrates the construction of the modulator including the materials, dimensions, and electrical contact vias. Fig. 2(b) gives a microscope image of a complete 11 µm diameter reflection modulator, including a signal contact and two ground contacts designed to minimize electrical parasitics. Fig. 2(c) gives a scanning electron microscope (SEM) image of the gold pad of an 11 µm diameter device bearing a plasmonic grating taken during the fabrication process. p-Si and HfO₂ are Si CMOS materials, but gold is not – gold was used in our experimental devices but could be replaced with a CMOS metal, *e.g.*, copper.

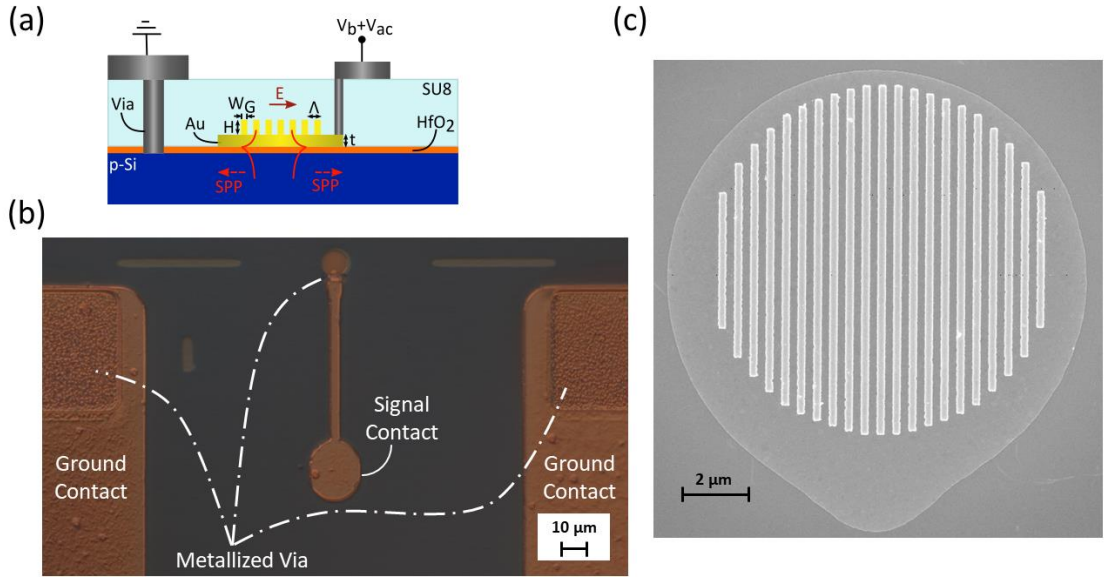

Figure 2. (a) Schematic of the modulator cross-section. (b) Optical microscope image of an 11 µm diameter modulator with ground-signal-ground contact pads. (c) Scanning electron microscope (SEM) image of an 11 µm diameter gold pad bearing a grating coupler (die O6) acquired during the fabrication process.

**Theoretical**

To study the performance of various designs, numerical calculations were performed in 2D for device diameters of 5, 11, and 17 µm, illuminated by a 5 µm diameter Gaussian beam aligned and focused on the plasmonic grating, in a scheme comparable to our experimental conditions. To account for the carrier refraction effect, a thin (1 nm) layer in the semiconductor adjacent to the oxide was assumed to have carrier density and refractive index perturbations (Δn = -0.5 and Δk = 0.1 for $\Delta N_a = 10^{20}$ cm$^{-3}$) associated with the applied AC voltage [30]. Simulations were done using a commercial finite element method (FEM) analysis tool and the structure was surrounded by a perfectly matched layer ending to scattering boundary conditions to avoid any artificial numerical back reflections. The gold and doped silicon permittivity were adopted from the literature [34]. The thickness and refractive index of the hafnia layer were considered as $H_{HfO_2}$ = 5 nm, and $n_{HfO_2}$ = 1.88 [35] respectively. The refractive index of the SU8 cover layer was assumed as $n_{SU8}$ = 1.57 [36]. A thin (1 nm) silicon native oxide layer was also added on the silicon substrate.

Fig. 3(a) shows the calculated reflectance of modulators of different diameters, designed at $\lambda_0$ = 1550 nm, with a grating pitch of Λ = 440 nm, ridge width of $W_G$ = 220 nm (50% duty cycle), and ridge thickness of H = 81 nm, in the absence of the thin perturbed layer (*i.e.*, the carrier density and refractive index of the silicon substrate adjacent to the oxide layer are unperturbed and set to bulk doped silicon). The plot reveals the main reflectance resonance along with a shoulder on the long-wavelength side for the 11 and 17 µm diameter devices. Comparing the transverse electric field at the main resonance wavelength and at its shoulder, as plotted in Figs. 3(d) and 3(e) for the 11 µm diameter device, suggests that the propagating surface plasmons are reflected from the edges of the circular plasmonic patch and interfere to create a standing wave along the patch which causes the appearance of another resonance in the reflection spectrum of the devices [37]. This secondary resonance can also appear if a tightly focused beam is incident on a device of greater area because it contains a broad spectrum of *k*-vectors, or if a device is illuminated by a plane wave or beam incident at an angle [38]. The latter can be observed in Fig. 4(c) (*cf.*, Experimental section).

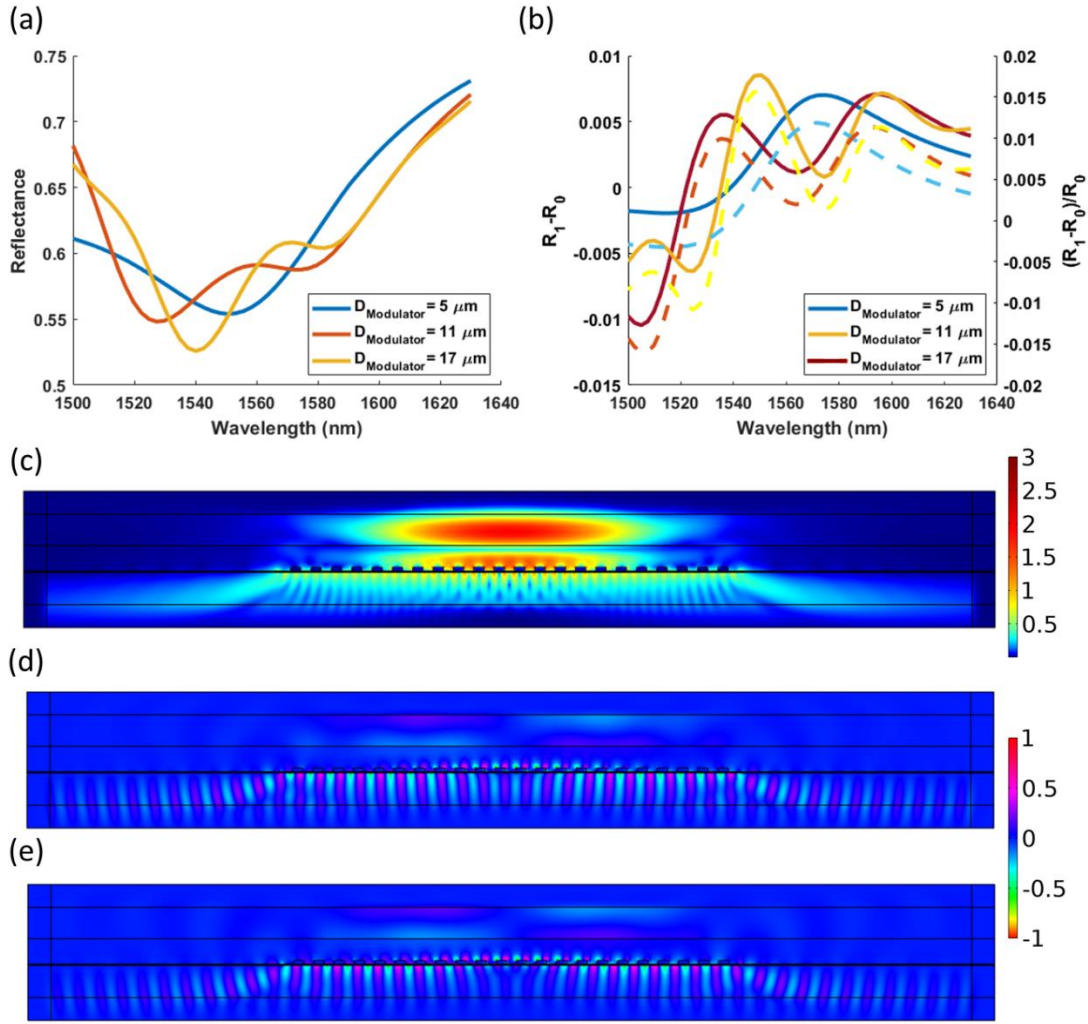

Figure 3. (a) Calculated reflectance response of modulators of different diameters. (b) Calculated reflectance modulation (solid curves) and normalized reflectance modulation (dashed curves) of devices in Part (a). (c) Total electric field magnitude (V/m) at λ = 1527 nm for the 11 µm diameter modulator. (d,e) Real part of transverse electric field, Re{$E_y$}, (V/m) at λ = 1527 nm and λ = 1574 nm. The incident field strength in all cases was set to 1 V/m.

Fig. 3(b) presents the reflectance modulation response produced by alternating between the unperturbed ($R_0$) and perturbed ($R_1$) cases, corresponding to unchanged carrier density with respect to the bulk silicon, and accumulated carriers in silicon adjacent to the oxide, respectively. It can be observed that the reflectance modulation has a broad optical bandwidth due to the broad nature of the plasmonic resonance, which peaks around the resonance wavelength where plasmons are most efficiently excited along the bottom metal surface bounded by the oxide and semiconductor regions of the device. Fig. 3(c) clearly reveals the excitation of SPPs along this bottom surface.

**Experimental**

Fig. 4(a) illustrates the optoelectronic measurement setup. The modulator chips were optically illuminated by a polarization maintaining (PM) tapered cone fiber (WT&T, model CL5) at λ = 1550 nm, with a focal spot diameter of 5 µm at a working distance of ~40 to 50 µm. The polarized output of a tunable diode laser (Photonetics Tunics Plus) was injected into this lensed fiber by first passing through a PM in-fiber optical circulator (OZ Optics).

The intensity-modulated beam reflected from the chip was then coupled back to the lensed fiber and directed out of the third port of the circulator. The modulated optical signal was injected into a fiber-coupled Mach-Zehnder interferometer modulator (MZIM) (JDSU OC-192) operating as a transparent patch cord during the measurements (undriven), or as an intensity modulator to calibrate the set-up. The modulated beam was then amplified using an Er-doped fiber amplifier (EDFA, JDSU AFC) with a fixed 15 dB gain, and attenuated to maintain the optical power to just below the 1 mW maximum input of our high-speed photoreceiver (Thorlabs PDA8GS). The beam was also filtered using a narrow band-pass in-fiber filter (OZ Optic TF) to reduce the amplified spontaneous emission (ASE) noise produced by the EDFA.

To drive the modulators electrically, the chips were contacted using a 50 Ω terminated ground-signal-ground (GSG) microwave probe (Picoprobe, model 40A). The probe was connected to a bias Tee network (Minicircuits ZBT-K283+) by a short RF cable (Minicircuits KBL-2FT-LOW+) to reduce loss and parasitic effects. The appropriate DC bias and AC drive voltages were then superimposed using the bias Tee. The chips were biased around an optimum DC voltage of $V_{Bias} = -1$ V which is close to the flat-band condition of the MOS structure ($V_{FB} = -0.8$ V) using a source meter unit (Keithley 2400). The output of a vector network analyzer (VNA HP8510B) was amplified by an RF amplifier (Minicircuits ZVA-02303HP+) to achieve an AC voltage of $V_{pp} \cong 4$ V, driving the MOS structure into deep accumulation in the negative half-cycle, then into depletion during the next half-cycle. This ensures maximum intensity modulation of the beam. The electrical modulation was recovered by detecting the intensity-modulated beam using a wideband photoreceiver, which was then used to drive an oscilloscope or redirected to the receiving port of the VNA for S21 measurements.

Before carrying out electro-optic measurements, the modulator chips were characterised optically and electrically. The electrical measurements consisted of acquiring capacitance-voltage (CV) characteristics. To do so, the electrical path presented in Fig. 4(a) was modified by removing the RF amplifier and substituting the terminated probe by an open-circuit probe (Picoprobe, model 40A) as sketched in the inset of Fig. 4(b). The VNA was then calibrated to the reference planes set by the tip of the probe by contacting elements on a calibration substrate (GGB industries CS-5). The reflection coefficient (S11) of the chip was then measured at different DC bias points keeping the frequency and the small AC voltage amplitude constant. The device capacitance was then extracted, and a CV characteristic obtained, as plotted in Fig. 4(b) for a 5 µm diameter device.

The optical response of a plasmonic grating patch was then obtained by measuring its reflectance response in a free-space telescope setup [30]. Fig. 4(c) gives an example response for a 28 µm diameter modulator measured under open-circuit conditions, normalized to the response of a mirror of similar diameter (a modulator without a plasmonic grating). Fig. 4(c) also plots the theoretical response (computed following the details given under theoretical) for a modulator of the same diameter, excited by a Gaussian beam of 15 µm waist diameter, following the conditions in our free space telescope setup and assuming a 3-degree misalignment with respect to normal incidence. The theoretical and measurement results match very well including the second resonance feature. The secondary resonance is not present if a loosely focused beam is incident on a device at normal incidence (*cf.*, Theoretical section). The resonance is slightly deeper in the calculations. This could be explained by deviations in material and/or structural properties between the theoretical model and the experimental structure. Furthermore, small translational misalignments along the optical beam path in the free-space setup can also affect the response of the device.

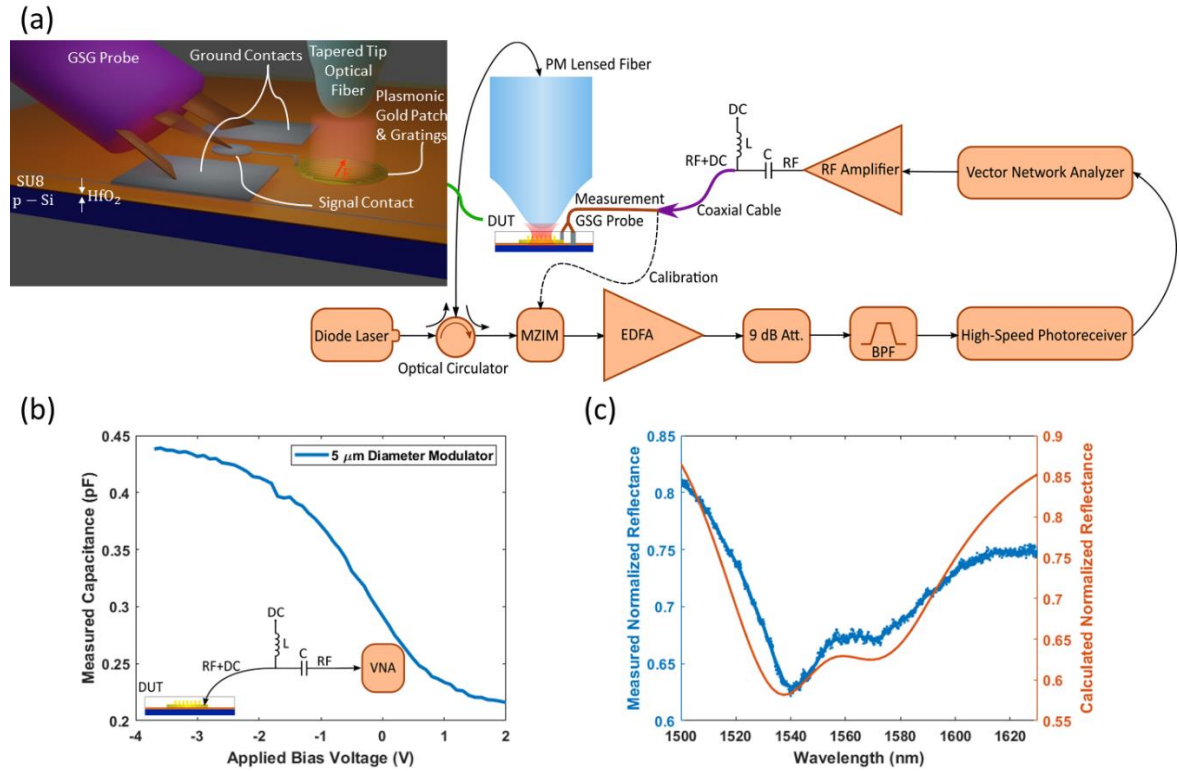

Figure 4. (a) Sketch of the calibration and measurement setup for the optoelectronic characterisation of modulators via S21 measurements. The sketch in inset presents a closer view of the device probed by a ground-signal-ground probe and illuminated by a tapered optical fiber. (b) Measured high frequency (500 MHz) CV characteristic of a 5 µm diameter modulator. Inset shows the simplified RF setup for S11 measurements. (c) Experimental and theoretical normalized reflectance response of a 28 µm diameter modulator (plotted on different scales).

The frequency response and equivalent circuit of modulators were determined by measuring the S11 response using open-circuit and 50 Ω terminated probes in the simplified setup presented in inset of Fig. 4(b), then extracting an equivalent circuit model from the measurements. Fig. 5(a) presents the response of a 5 µm diameter device measured using the open circuit probe, at a DC bias point of $V_{\text{Bias}} = 2$ V (which produces the lowest capacitance and broadest response), and Fig. 5(b) shows the results obtained using the 50 Ω terminated probe on the same device at the same bias point. For the measurements using the open-circuit probe, the VNA and probe were calibrated as described in the previous section (as for CV measurements), but for the measurements with the terminated probe, the VNA reference plane was set by calibration to the end of the RF cable (before the probe) and an electrical delay was added to account for the electrical length of the probe. The measurement of Fig. 5(b) yields an electrical 3 dB bandwidth of BW ≅ 22 GHz.

Fig. 5(c) presents the S11 responses of a 10 µm diameter modulator obtained using the 50 Ω terminated probe at bias points of $V_{Bias} = -1, 0,$ and 1 V, from which 3 dB electrical bandwidths of BW = 5, 7 and 9 GHz are observed, respectively. It can be concluded that by moving the bias point towards the inversion regime (positive bias voltages), the 3 dB bandwidth of the modulators increases due to a decrease in the device capacitance, as also supported by the CV characteristics of Fig. 4(b). However, to achieve the best modulation depth, the devices must be driven into deep accumulation to benefit from the carrier refraction effect. As a result, a DC bias point of $V_{Bias} = -1$ V was chosen for the forthcoming electro-optic measurements.

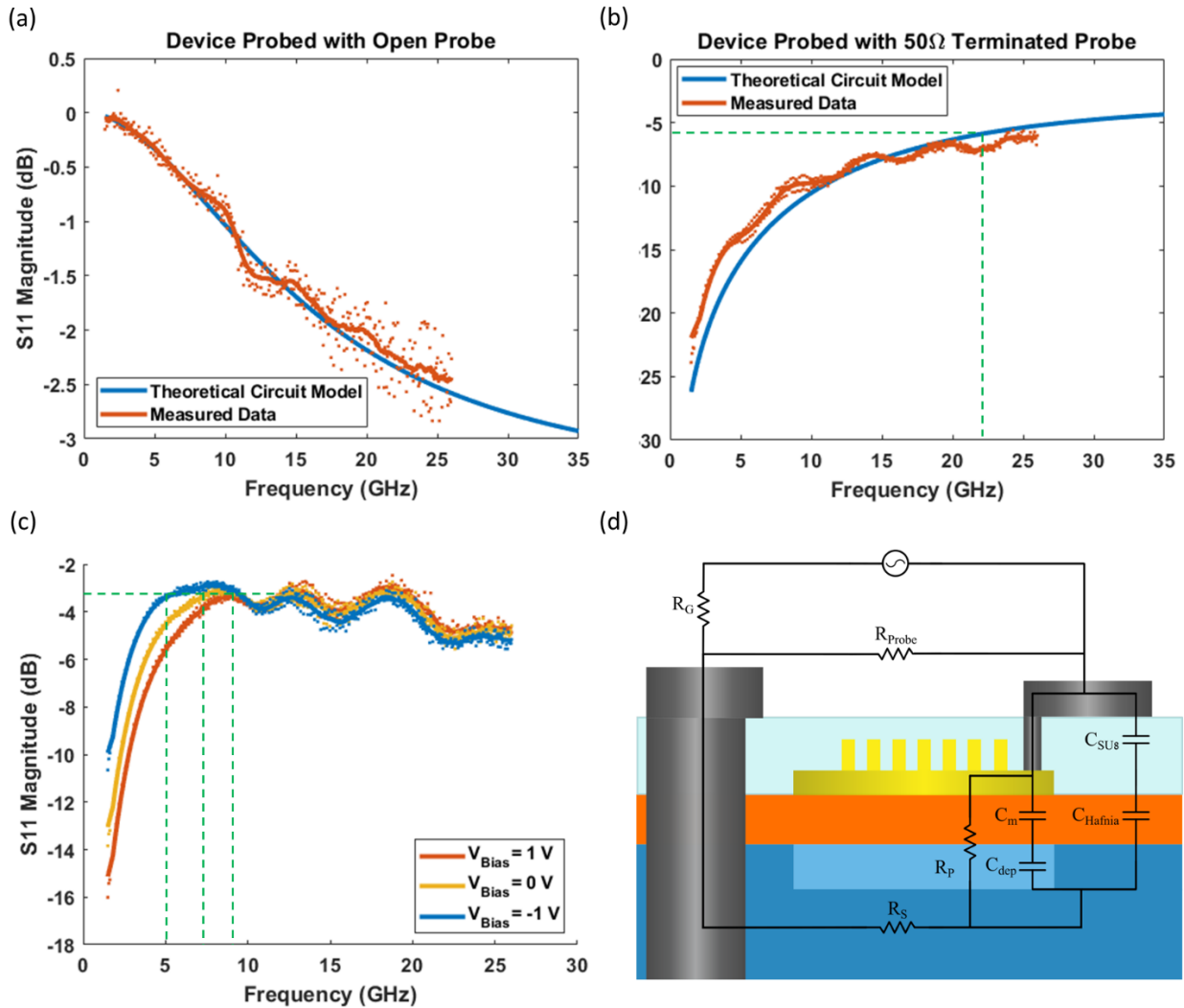

**Figure 5.** (a) S11 experimental and theoretical (equivalent circuit) frequency responses of a 5 µm diameter modulator probed with an open circuit probe and biased at $V_{Bias} = 2\ V$. (b) Experimental and theoretical S11 frequency responses of the same modulator as in Part (a) ($V_{Bias} = 2\ V$) but probed with a 50 Ω terminated probe. (c) S11 frequency responses of a 10 µm diameter modulator measured with a 50 Ω terminated probe, at bias voltages of $V_{Bias}$ = -1, 0 and 1 V. A moving average of window length 20 is superimposed on the measured data points in Parts (a) to (c). (d) Equivalent circuit for which the frequency response is also plotted in Parts (a) and (b), superimposed on the modulator structure of Fig. 2(a), showing the origin of each circuit element. The sketch is not to scale and the depletion layer in Si is shown in light blue. The generator is also shown connected to the equivalent circuit.

The equivalent electrical circuit of the modulators is shown in inset to Fig. 5(a) and in Fig. 5(d) which illustrates the origin of the elements relative to the device cross-section. The MOS structure has a depletion capacitance which is DC bias dependent, modelled as a parallel plate capacitor [39]:

$$C_{dep} = \frac{\epsilon_0 \epsilon_{Si} A_m}{x_d} \quad (1)$$

where $\epsilon_0$ and $\epsilon_{Si}$ = 11.9 are the vacuum permittivity and DC relative permittivity of the silicon substrate, respectively, $A_m$ is the area of the modulator, and $x_d$ is the bias dependant depletion width, given by [39]:

$$x_d = \sqrt{\frac{2\epsilon_0 \epsilon_{Si} \phi_S}{qN_A}} \quad (2)$$

where $N_A$ is the equilibrium density of holes in bulk Si and q is the electron charge. $\Phi_S$ is the surface potential of the MOS structure at the applied bias. Assuming we have reached strong inversion at the bias voltage of $V_{Bias}$ = 2 V, $\phi_S \cong 2\phi_{Bp}$, where $\phi_{Bp}$ is the potential difference between the Fermi and intrinsic levels of the semiconductor [39]. A bias voltage of $V_{Bias}$ = 2 V yields $x_d$ = 7 nm and $C_{dep}$ = 0.4 pF for the 5 μm diameter device. The depletion capacitance appears in series with the oxide capacitance $C_m$ which is calculated using a parallel plate capacitor model similar to Eq. (1), but for a 5.5 nm thick $HfO_2$ layer as the insulating oxide with $\epsilon_{HfO_2} \cong 10.7$ (relative permittivity) [31]. An oxide capacitance of $C_m$ = 0.44 pF is obtained for the 5 μm diameter device, which is very close to the capacitance measured in the accumulation regime in Fig. 4(b). Parasitic capacitances between the signal pads and the grounded silicon substrate also appear. These capacitances each include two capacitive contributions in series due to the SU8 and $HfO_2$ layers separating the pads from ground, calculated as $C_{SU8}$ = 3 fF ($\epsilon_{SU8}$ = 4 is the relative permittivity of SU8 [40] [41]) and $C_{HfO_2}$ = 0.86 pF. The total capacitance of the equivalent circuit for the 5 μm diameter device is then $C_T \cong 0.21$ pF which is close to the minimum capacitance measured in the inversion region of Fig. 4(b). The parasitic series resistance of the modulators was assumed to originate from the metalized vias and was determined by measuring the resistance of via pairs in 10 interconnected rows of vias, yielding a value of $R_S$ = 9 Ω. The shunt resistance of the devices was extracted from LCR measurements and found to be least $R_P$ = 1 MΩ. Using this equivalent circuit, the calculated S11 frequency response of the 5 μm diameter modulator is plotted in Fig. 5(a) for the case of probing with the open-circuit probe ($R_{Probe} = \infty$) and in Fig. 5(b) for the terminated probe ($R_{Probe}$ = 50 Ω) - excellent agreement with the measurements is noted in both cases. The electrical bandwidth of the modulators can be calculated as $BW = (2\pi R_T C_T)^{-1}$ where:

$$C_T = \frac{C_m C_{dep}}{C_m + C_{dep}} + \frac{C_{SU8} C_{Hafnia}}{C_{SU8} + C_{Hafnia}} = 0.21 \text{ pF} \quad (3)$$

and

$$R_T = (R_S + (R_{Probe} \parallel R_g)) \parallel R_p = 34 \text{ Ω} \quad (4)$$

are the total capacitance and resistance of the equivalent circuit. Using these values, the BW works out to 22 GHz when probed with the 50 Ω terminated probe, in agreement with the measurement (Fig. 5(b)). It's worth noting that probing the modulator chips with a terminated 50 Ω probe, which is impedance matched to an RF system of 50 Ω characteristic impedance, broadens the working bandwidth by reducing the total resistance of the circuit.

Fig. 6(a) shows the measured electro-optic transmission coefficient (S21, normalized) of two modulator chips of 10 and 5 µm diameter, each at two bias points. The electro-optic calibration path for S21 measurements is as sketched in Fig. 4(a), and includes the modulator under test, and all RF and optical components in the measurement path except for the microwave GSG probe. To calibrate the set-up, the MZIM was excited with similar RF ($V_{pp} = 4$ V) and optical (0.4 mW) signals and the measured S21 response of the chain taken as the calibrated response in the VNA. Thus, all subsequent modulator measurements are relative to the MZIM response.

From Fig. 6(a), a 3 dB electro-optic bandwidth of BW $\cong$ 5 GHz is evident from the S21 frequency response of the 11 µm device at $V_{Bias} = -1$ V, in agreement with the bandwidth obtained from the S11 measurements presented in Fig. 5(c) for the same bias point. For a bias voltage closer to the inversion regime, the 3 dB bandwidth of the same device increases, and the response becomes flatter. This is evident from the S21 curve presented in Fig. 6(a) for $V_{Bias} = 0$ V and agrees with the BW of 7 GHz extracted from Fig. 5(c). However, the modulation depth decreases as the bias point moves towards the inversion regime, which reduces the magnitude of S21. A similar behaviour is observed for the 5 µm diameter device at the same two bias points. The measured electro-optic S21 frequency response of the 5 µm modulator was limited by the bandwidth of our photoreceiver (8.5 GHz), but its 3 dB electrical bandwidth is ~22 GHz as measured previously (Fig. 5(b)). The larger bandwidth of this device results in improved flatness of the S21 frequency response compared to that of the 11 µm diameter device.

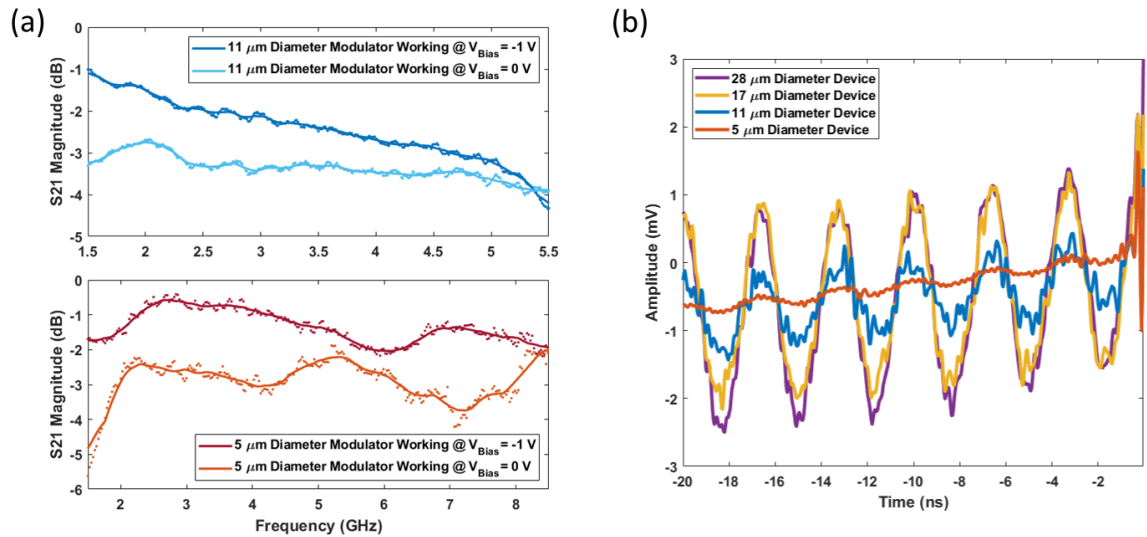

**Figure 6.** (a) Measured normalized electro-optic transmission coefficient magnitude (|S21|) for 11 and 5 µm diameter modulators at different bias voltages. A moving average of window length 20 is superposed on the measured data points. (b) Low frequency (300 MHz) modulated signal detected by a photoreceiver and measured on an oscilloscope for modulators of different diameters, (averaged over 50 waveforms).

Fig. 6(b) shows the low-frequency modulation signal from devices of different diameters, measured on an oscilloscope (Tektronics TDS694C), and averaged over 50 waveforms to reduce noise. For the low frequency measurements, a different RF power amplifier (mini-circuits TVA-11-422) and bias-tee circuit (mini-circuits ZX85-12G-S+) were used. The modulation signal amplitude decreases with the device diameter under similar experimental conditions. This can be explained by the fact that alignment of the beam with a spot diameter of 5 µm at the output of our lensed fiber, is more challenging as the device diameter shrinks. The area of the 5 µm diameter modulator does not overlap with the entire area of the incident beam, so part of the latter is scattered.

The insertion loss of the modulators was determined by comparing its output optical power at the third port of the circulator relative to that of a broadband perfect mirror placed at the same location as the modulator in the set-up. The insertion loss of the modulators depends on the device size and varies from $\text{IL} \cong 4.2$ dB to $\text{IL} \cong 2$ dB for the 5 and 28 µm diameter modulators, respectively. These insertion loss values are relative to the incident free space optical beam and represent those of the modulator alone (without the effect of the ancillary components in the set-up).

The normalized reflectance modulation, also referred to as the modulation depth (MD), is:

$$\text{MD} = \left|\frac{R_1 - R_0}{R_0}\right| = \left|\frac{P_{max} - P_{min}}{P_{min}}\right| = \left|\frac{V_{max} - V_{min}}{V_{min}}\right| \tag{5}$$

where the third equality emerges by recognizing that the voltage measured on the oscilloscope is proportional to the power output from the modulator ($P_{max}$, $P_{min}$). Using the traces of Fig. 6(b) we find MD = 0.12%, 1.3%, 1.9% and 2.3% for the 5, 11, 17, and 28 µm diameter modulators, respectively, which are in good agreement with our numerical results plotted on Fig. 3(c). The modulation depth is low but should be adequate for wafer probing purposes assuming high-quality receiver optoelectronics. The modulation depth can be improved by choosing, *e.g.*, epsilon near zero materials (transparent conductive oxides) [21] [24], however, adherence to CMOS material sets represents a tight constraint that should be respected for a wafer testing application.

**Conclusion**

A MOS based plasmonic modulator concept of small footprint (5 µm diameter), wide optical (100 nm) and electrical (22 GHz) working bandwidths, and low insertion loss (4 dB) was proposed and demonstrated. The modulation depth of the modulators is about 2%, which should be adequate for the device to operate as a data output modulator in non-contact wafer-level testing. Wafer testing impacts the Si CMOS wafer manufacturing industry and has significant economic consequences. The electro-optic modulator concept proposed and investigated in this paper meets the main requirements for a non-contact solution to this problem. By changing the metal from Au to, *e.g.*, Cu the modulators could be monolithically integrated with Si CMOS electronics, consuming a diameter of about 5 µm on wafer, and they would scale without crosstalk.

**Acknowledgments**

The authors are grateful to Anthony Olivieri and Saba Siadat Mousavi for device fabrication, and to Jan Hoppe for fruitful discussions.

**Funding resources**

Natural Sciences and Engineering Research Council of Canada (210396)